\documentclass[showpacs,amssymb,floatfix]{revtex4}
\usepackage{graphicx}
\usepackage{dcolumn}
\usepackage{graphics}
\usepackage{epstopdf}
\usepackage{pdfpages}
\usepackage{bm}
\begin{document}
\title{Bloch bound states in the radiation continuum in a periodic array of dielectric rods}
\author{Evgeny N. Bulgakov and Almas F. Sadreev}
\address{L.V. Kirensky Institute of Physics, 660036 Krasnoyarsk,
Russia}
\date{\today}
\begin{abstract}
We consider an infinite periodic  array of dielectric rods in vacuum with the aim to demonstrate
three types of a Bloch bound states in the continuum (BSC),
symmetry protected with a zero Bloch vector, embedded into
one diffraction channel with nonzero Bloch vector,
and embedded into two and three diffraction
channels. The first and second types of the BSC exist in a wide range of material parameters of the rods,
while the third occurs only at a specific value of the radius of the rods.
We show that the second type supports the power flux along the array.
In order to find BSC we put forward an approach based on the expansion over the Hankel
functions. We show how the BSC reveals itself in the scattering function when the singular BSC point is
approached along a specific path in the parametric space.
\end{abstract}
\pacs{42.25.Bs,42.65.Jx,03.65.Nk,42.25.Fx}

\maketitle
\section{Introduction}

In 1929, von Neumann and Wigner \cite{neumann} predicted the
existence of discrete solutions of the single-particle
Schr\"odinger equation embedded in the continuum of positive
energy states, bound states in the continuum (BSC). Their analysis examined by Stillinger and Herrick
\cite{stillinger} for a long time was regarded as mathematical curiosity
because of certain spatially oscillating central symmetric
potentials. That situation cardinally changed when Friedrich and Wintgen \cite{friedrich}
formulated a generic two-level Fano-Anderson model and derived a condition for
BSC  as a resonant state whose width tends
to zero as at least one physical parameter varies continuously
(see, also \cite{volya,guevara,SBR}). Nowadays one can see a rapid growth in the amount of
publications, both theoretical and experimental devoted to the BSCs in different physical systems
\cite{Shipman,Porter,photonic,Shabanov,Moiseyev,Prod,FPR,Lepetit,Gonzalez,anneal,Segev,Longhi,
Kivshar,Wei1,Wei,Yang,robust}.

The BSC can be classified by the mechanism responsible for the localization of waves \cite{Gonzalez,anneal}.
The most obvious mechanism is related to the symmetry \cite{Moiseyev,Shipman} when
the continuum states  and the BSCs have different symmetries.
The second mechanism is a Fabry-Perot resonator with the wave trapped  between  the mirrors as
the distance between the mirrors is tuned. That mechanism was
used in the system of quantum dots coupled by a wire \cite{Kim,RS} as well as in photonics
\cite{FanPRB,photonic,Shabanov,Ndangali,FPR}. The above mentioned types of the BSC were extensively studied in different
photonics structures and experimentally observed
\cite{Lepetit,Segev,Longhi,Kivshar,Wei1}.
The third mechanism is full destructive interference of two resonances
which was originally put forward by Friedrich and Wintgen \cite{friedrich} and later developed in Refs.
\cite{SBR,robust}.
It should be noted however that experimental observation of this type of BSC  is difficult because it is necessary
to vary material parameters \cite{Wei1}.
Fourth, at some value of parameters the coupling of a bound state of the closed system
with the continuum channel can turn to zero accidentally to give rise to the accidental BSC \cite{anneal}.
We speculate that this mechanism underlies the robust BSC
observed in photonic crystal slab \cite{Wei,Yang}.
In the present paper we consider a one-dimensional periodic array of GaAs cylindrical rods.
We show that the array is capable to support multiple symmetry protected Bloch BSC.
The Bloch BSC with nonzero wave vector  supports power flux.
We also show that there are Bloch BSCs embedded into two and three continua.
\section{Basic equations for EM wave scattering by periodic array of rods}
\begin{figure}[ht]
\includegraphics[width=7cm,clip=]{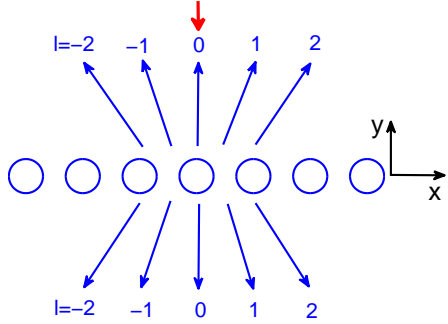}
\caption{(Color online) Cross section of a periodic array of parallel infinitely long rods illuminated by a plane
wave (thick red arrow). The wave can be transmitted or reflected to discrete
diffraction channels enumerated by integer $l$ shown by blue arrows.} \label{fig1}
\end{figure}

The system under consideration is infinitely long array of GaAs rods. The rods are infinitely long in the z direction,
parallel to each other, and periodically spaced with the distance h along the x-axis on the x-y plane
as shown in Fig. \ref{fig1}. In what follows we take $h=1$.
We consider the scattering of TM electromagnetic waves by this array.
The scattering of  plane waves  by  cylinders  has  been  the  subject of  many
investigations long time ago.  Most  of difficulties in  this  connection  is caused by
multiple scattering by cylinders \cite{Twersky}. For the case of two cylinders \cite{Young}
and infinite periodic row of cylinders \cite{Linton,Yasumoto} the problem was considered
to bring it to a simple and tractable formulation.
In this section we present the basic equations of that theory for the reader's convenience based on Ref. \cite{Yasumoto}.

Assume that the periodic array of rods is illuminated by a plane wave
$$\psi_{inc}(x,y)=\sqrt{\frac{2}{|k_y|}}e^{i(k_xx+k_yy)}$$ where
$\psi$ is the electric field directed along the rods,
$k_x=-k_0\cos\varphi_i, k_y=-k_0\sin\varphi_i$, $k_0=\omega/c$ and  $\varphi_i$ defines the angle of incidence.
The plane wave  can be expanded over the Bessel functions
\begin{equation}
\label{eq1}
  \psi_{inc}(r,\varphi)=\sqrt{\frac{2}{|k_y|}}\sum_m(-i)^me^{im(\varphi-\varphi_i)}J_m(k_0r)=
  \sum_m\psi_{inc,m}e^{im\varphi}J_m(k_0r),
\end{equation}
where
\begin{equation}\label{columns}
\psi_{inc,m}=\sqrt{\frac{2}{|k_y|}}{(-i)^me^{-im\varphi_i}}=\sqrt{\frac{2}{|k_y|}}
\left(\frac{ik_x+k_y}{k_0}\right)^m.
\end{equation}
The scattered wave outside the rod is given by the Hankel functions
\begin{equation}
\label{eq2}
\Psi_s(r,\varphi)=\sqrt{\frac{2}{|k_y|}}\sum_{m}a_m\exp{(im\varphi)}H_m^{(1)}(k_0r)
\end{equation}
The relation between incident  and scattered waves
is given by
\begin{equation}
a_m=\sum_nT_{mn}\psi_{inc,n},
\end{equation}
where the transition matrix has the following form
\begin{equation}
\label{eq3}
T_{mn}=\delta_{mn}\frac{\sqrt{\epsilon}J^{'}_{m}(qR)J_m(k_0R)-J^{'}_{m}(k_0R)J_m(qR)}
{H^{(1)'}_{m}(k_0R)J_m(qR)-\sqrt{\epsilon}J^{'}_{m}(qR)H^{(1)}_m(k_0R)},
\end{equation}
$q=\sqrt{\epsilon}k_0$, and $\epsilon$ is the permittivity of the rod of radius $R$.
The scattering function is a sum of waves scattered from each rod:
\begin{equation}\label{scatt}
\Psi_s(x,y)=\sqrt{\frac{2}{|k_y|}}\sum_j\sum_ma_{m,j}\exp{(im\varphi_j)}H_m^{(1)}(k_0r_j)
\end{equation}
where $(r_j, \varphi_j)$ is the local polar coordinate system whose origin is located in the
center of the j-th rod and $a_{m,j}$ denotes the amplitude of the field scattered
from the j-th rod. The periodicity of the structure requires that scattered field satisfies
the  Bloch theorem:
\begin{equation}
\label{eq5}
\Psi_s(x+1,y)=\exp{(ik_x)}\Psi_s(x,y)
\end{equation}
with $k_x$ as the Bloch vector. Substituting Eq. (\ref{eq5}) into Eq. (\ref{scatt}) we have
\begin{equation}
\label{eq6}
\Psi_s(x,y)=\sqrt{\frac{2}{|k_y|}}\sum_j\sum_m\exp (ik_xj)a_m\exp (im\varphi_j)H_m^{(1)}(k_0r_j),
\end{equation}
where $a_m=a_{m,0}$.
The total wave function therefore is $\Psi=\Psi_{inc}+\Psi_s$.
Consequently Eq. (\ref{eq5})
defines the wave vectors of diffraction channels \cite{Yasumoto, Ndangali}
\begin{equation}\label{kyl}
    k_{y,l}=\sqrt{k_0^2-k_{x,l}^2},  ~~k_{x,l}=k_x+2\pi l, l=0, \pm 1, \pm 2, \ldots.
\end{equation}
The infinite periodic array
of rods scatters the wave only into a number of directions defined by wave vectors $k_{x,l}, k_{y,l}$.
In the other words, the system under consideration supports only finite number of continua or diffraction channels
notified by integer $l$ in Eq. (\ref{kyl}) for which $k_{y,l}$ is real. The diffraction channel waves are given by
\begin{equation}\label{channels}
    \psi_l^{r,t}=\sqrt{\frac{2}{k_{y,l}}}e^{ik_{x,l}x\pm ik_{y,l}y}.
\end{equation}
Here indexes  $t,r$ imply the transmitted and reflected plane waves
which are schematically shown in Fig. \ref{fig1} by blue slim arrows.

Following Ref. \cite{Yasumoto} we introduce the aggregate matrix $\widehat{A}$
which relates the amplitudes $\mathbf{\Psi}=\{\psi_{inc,m}\}$
in the expansion of the incident plane wave (\ref{eq1}) to the amplitudes
$\mathbf{a}=\{a_m\}$ of scattered wave (\ref{eq6})
\begin{equation}\label{Svec}
\mathbf{a}=\widehat{A}\mathbf{\Psi}_{inc}.
\end{equation}
That matrix could be considered as an analogue of the scattering matrix however there is important difference.
The last connects the scattering channels notified by $l$ with the incident wave while the former
connects the amplitudes of the Hankel functions in the scattered wave with the incident wave.

For the infinite periodic array of rods multiple scattering events can be summated
as follows \cite{Yasumoto}
\begin{equation}\label{S1}
    \widehat{A}=(1-\widehat{T}\widehat{L})^{-1}\widehat{T}
\end{equation}
where $\widehat{T}$-matrix is given by Eq. (\ref{eq3}) and
\begin{equation}
\label{L}
L_{n-m}=\sum_{l=1}^{\infty}H^{(1)}_{n-m}(lk_0h)[e^{ik_xl}
+(-1)^{n-m}e^{-ik_xl)}].
\end{equation}

Let us introduce components of waves transmitted and reflected in the $l$-th diffraction channel
\begin{equation}\label{vtr}
    \mathbf{v}_l^{(t,r)}=
    \sqrt{\frac{2}{k_{y,l}}}\left\{\left(\frac{ik_{x,l} \mp k_{y,l}}{k_0}\right)^m\right\}
\end{equation}
in terms of which we rewrite Eq. (\ref{Svec}) in accordance to Eq. (\ref{columns})
\begin{equation}
\label{Svect}
\mathbf{a}=\widehat{A}\mathbf{v}_0^{(t)}.
\end{equation}
Then we have \cite{Yasumoto}
\begin{equation}\label{rt}
    r_l= [\mathbf{v}_l^{(r)}]^{+}\mathbf{a}, ~~ t_l=\delta_{l,0}+[\mathbf{v}_l^{(t)}]^{+}\mathbf{a},
\end{equation}
which are the reflection and transmission amplitudes respectively for open channels.
The scattering function (\ref{scatt}) can be written as follows \cite{Yasumoto}
\begin{equation}
\label{scattering}
\Psi_s(x,y)=\left\{\begin{array}{cc}
\sum_lr_l\psi_l^{(r)}(x,y) & \mbox{if $y>0$} \\ \sum_lt_l\psi_l^{(t)}(x,y) & \mbox{if $y<0$}
\end{array}\right.
\end{equation}
where both open and closed channels are summated.

\section{Calculation of  bound states in the continuum}
Typically, a plane wave is scattered by the array of rods into
diffraction channels according to Eq. (\ref{Svect}). However there could be a unique case
when the matrix $\widehat{A}$ in Eq. (\ref{Svect}) is singular and the solution for the scattering wave exists irrespective to
the incident wave amplitude $\mathbf{v}_0^{(t)}$. As will be shown below this unique solution  is decoupled
from the diffraction channels and localized nearby the array.
This solution is the bound state with frequency embedded in the diffraction continua, i.e., a BSC.
In this section we adapt the approach of the effective non-Hermitian Hamiltonian \cite{ring,PRA}
for calculation of BSC in the periodic array of rods.

As it could be seen from Eq. (\ref{Svect}) the aggregate matrix $\widehat{A}$ must be singular for the BSC to exist.
This matrix is not hermitian, and
similar to the approach of the effective Hamiltonian \cite{SR,Ingrid} it is fruitful to introduce the biorthogonal basis
of the eigenvectors  of the matrix
\begin{equation}\label{Lambda}
\widehat{\Lambda}=\widehat{A}^{-1}=\widehat{T}^{-1}-\widehat{L}
\end{equation}
as follows
\begin{equation}
\label{bio}
\widehat{\Lambda}\mathbf{x}_f=\lambda_f\mathbf{x}_f,
~~\widehat{\Lambda}^{+}\mathbf{y}_f=\lambda_f^{*}\mathbf{y}_f,
~~\mathbf{y}_f^{+}\mathbf{x}_g=\delta_{fg}.
\end{equation}
One can show that
\begin{equation}\label{bio1}
    \sum_f\mathbf{x}_f\mathbf{y}_f^{+}=1,
    ~~\widehat{\Lambda}^{-1}=\sum_f\frac{\mathbf{x}_f\mathbf{y}_f^{+}}{\lambda_f}.
\end{equation}
Then by the use of Eqs. (\ref{bio}) and (\ref{bio1}) the scattering state in Eq. (\ref{Svect}) can be expanded
over the biorthogonal basis as follows
\begin{equation}\label{scat}
    \mathbf{a}=\sum_f\frac{(\mathbf{y}_f^{+}\mathbf{v}_0^{(t)})\mathbf{x}_f}{\lambda_f}.
     \end{equation}

By substituting Eq. (\ref{scat}) into (\ref{rt}) we obtain the scattering amplitudes
in the following form
\begin{equation}\label{rltl}
    r_l=\sum_f\frac{W_{f,l}^{(r)}\widetilde{W}_{f,0}^{(t)}}{\lambda_f}, ~\
t_l=\delta_{l,0}+\sum_f\frac{W_{f,l}^{(t)}\widetilde{W}_{f,0}^{(t)}}{\lambda_f}
\end{equation}
where we introduced the coupling constants between the $f$-th eigenvector of the matrix $\Lambda$ and the
$l$-th diffraction channel vector $\mathbf{v}_0^{(t)}$
\begin{equation}\label{Wnl}
    W_{f,l}^{(r,t)}=[\mathbf{v}_l^{(r,t)}]^{+}\mathbf{x}_f,
    ~~\widetilde{W}_{f,0}^{(t)}=\mathbf{y}_f^{+}\mathbf{v}_l^{(t)}.
\end{equation}
Eqs. (\ref{rltl}) are similar to those derived in Ref. \cite{SR}.

The BSC occurs when, at least, one of complex eigenvalues $\lambda$ equals zero
\begin{equation}\label{BSC}
    \widehat{\Lambda}\mathbf{\Phi}_{BSC}=0
\end{equation}
Therefore, the BSC is in fact a null eigenvector.
Next, Eq.(\ref{rltl}) shows that for the scattering amplitudes to remain finite we have to imply
the condition
\begin{equation}\label{null}
    [\mathbf{v}_l^{r,t}]^{+}\mathbf{\Phi}_{BSC}=0.
\end{equation}
This equation shows that the BSC is decoupled from all open diffraction channels and therefore does not
leak into the radiation continuum.

\section{Symmetry protected BSC}
In what follows we take the dielectric constant of the rods
$\epsilon=12$ (GaAs rods). We briefly describe numerical procedure of calculation of the  BSC function.
At the first step the eigenvalue problem
for the matrix $\Lambda$ defined by Eqs. (\ref{L}) and (\ref{Lambda}) is solving. Numerically we
truncate  the rank of
the matrix $\Lambda$ by $|m|\leq 10$ to ensure sufficient accuracy. Then the eigen vector
components $a_m$ are substituted into Eq. (\ref{rt}) and after into Eq. (\ref{scattering})
to find the BSC wave function outside the rods. A summation over $l$
includes only closed channels, i.e. evanescent modes.
The BSC function inside the rods is to be expanded over the Bessel functions where the coefficients
are calculated by means of the matching of the outside and inside functions on the boundary of rods.

\begin{figure}[ht]
\includegraphics[width=7cm,clip=]{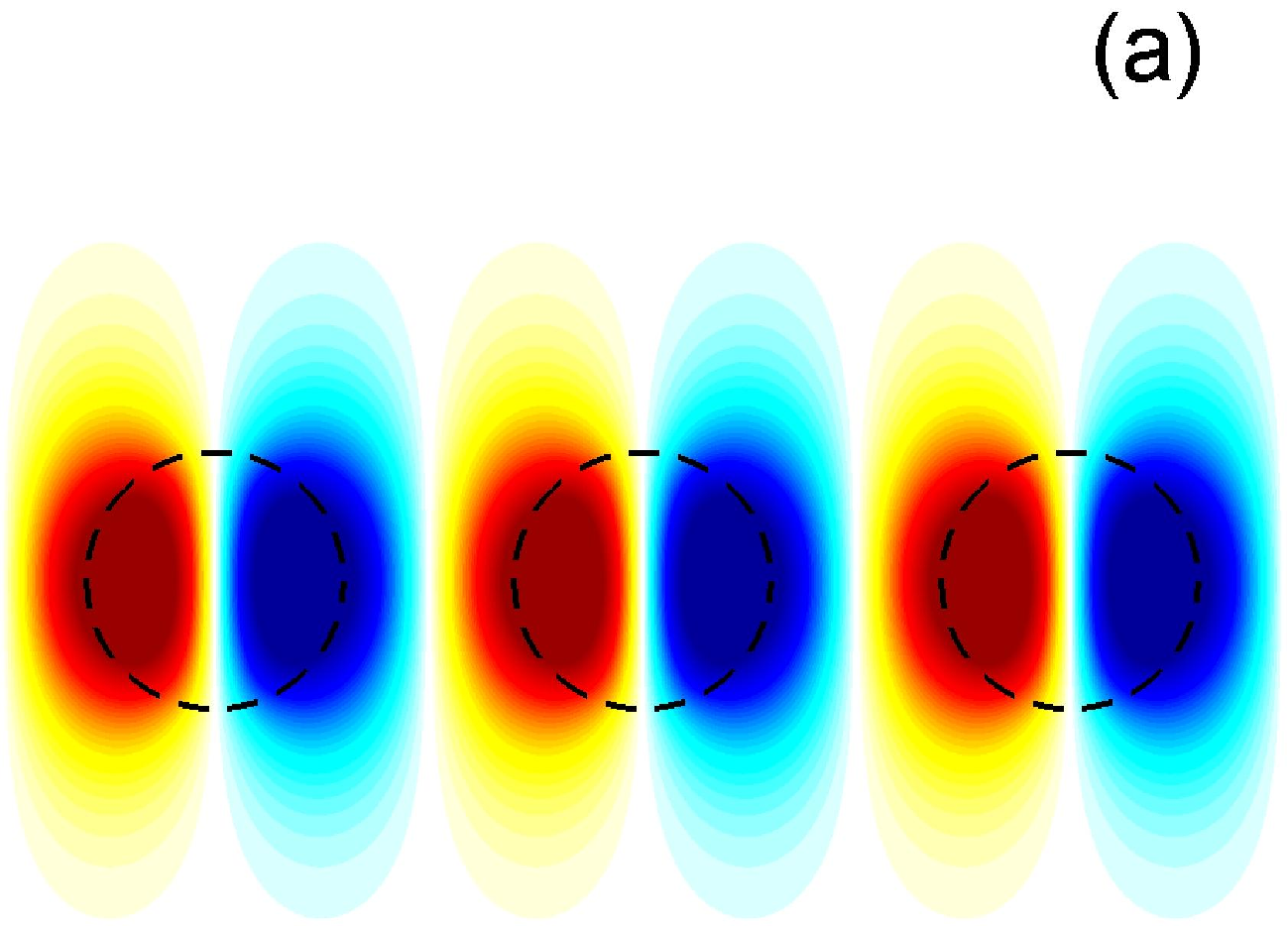}
\includegraphics[width=7cm,clip=]{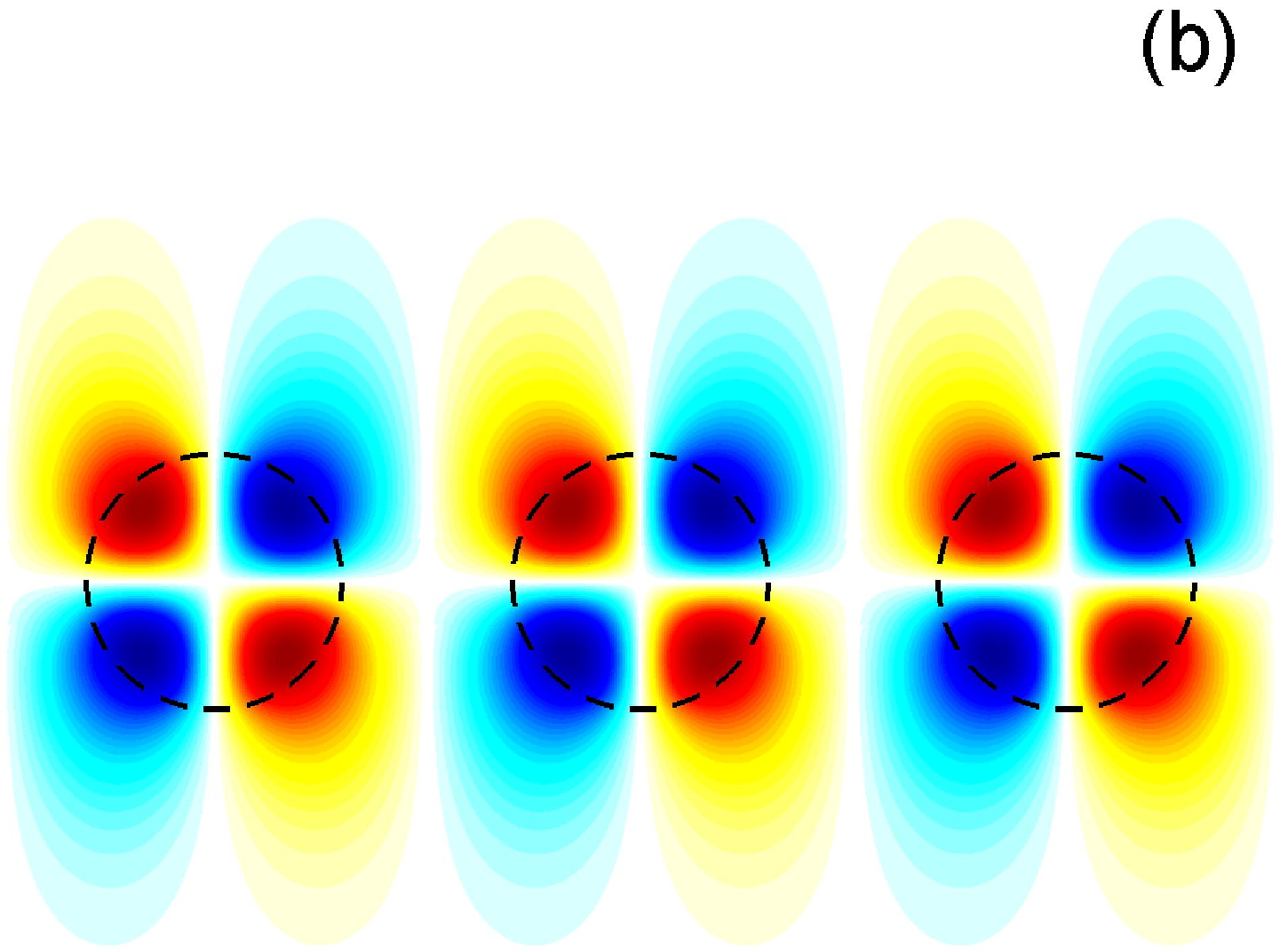}
\includegraphics[width=7cm,clip=]{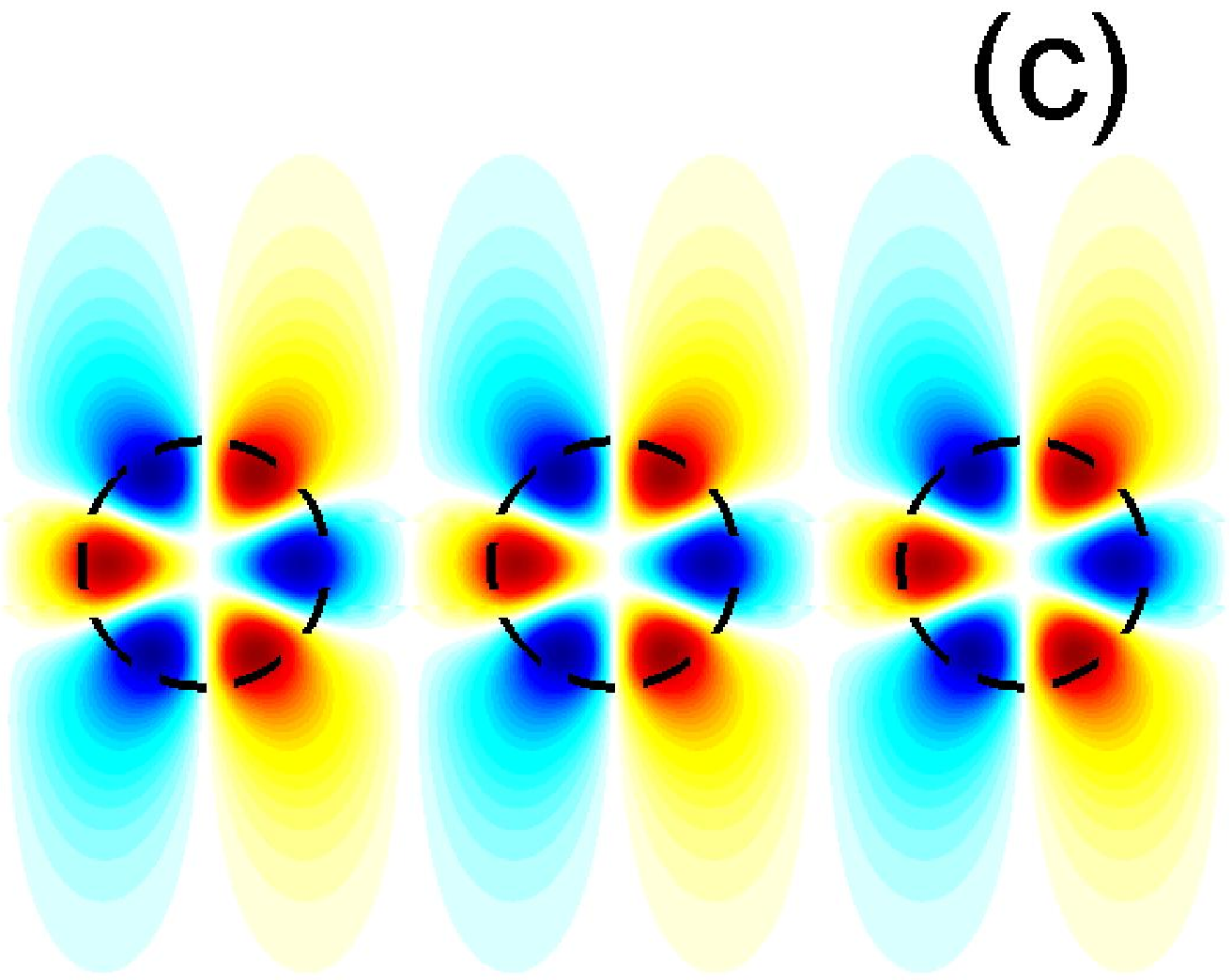}
\includegraphics[width=7cm,clip=]{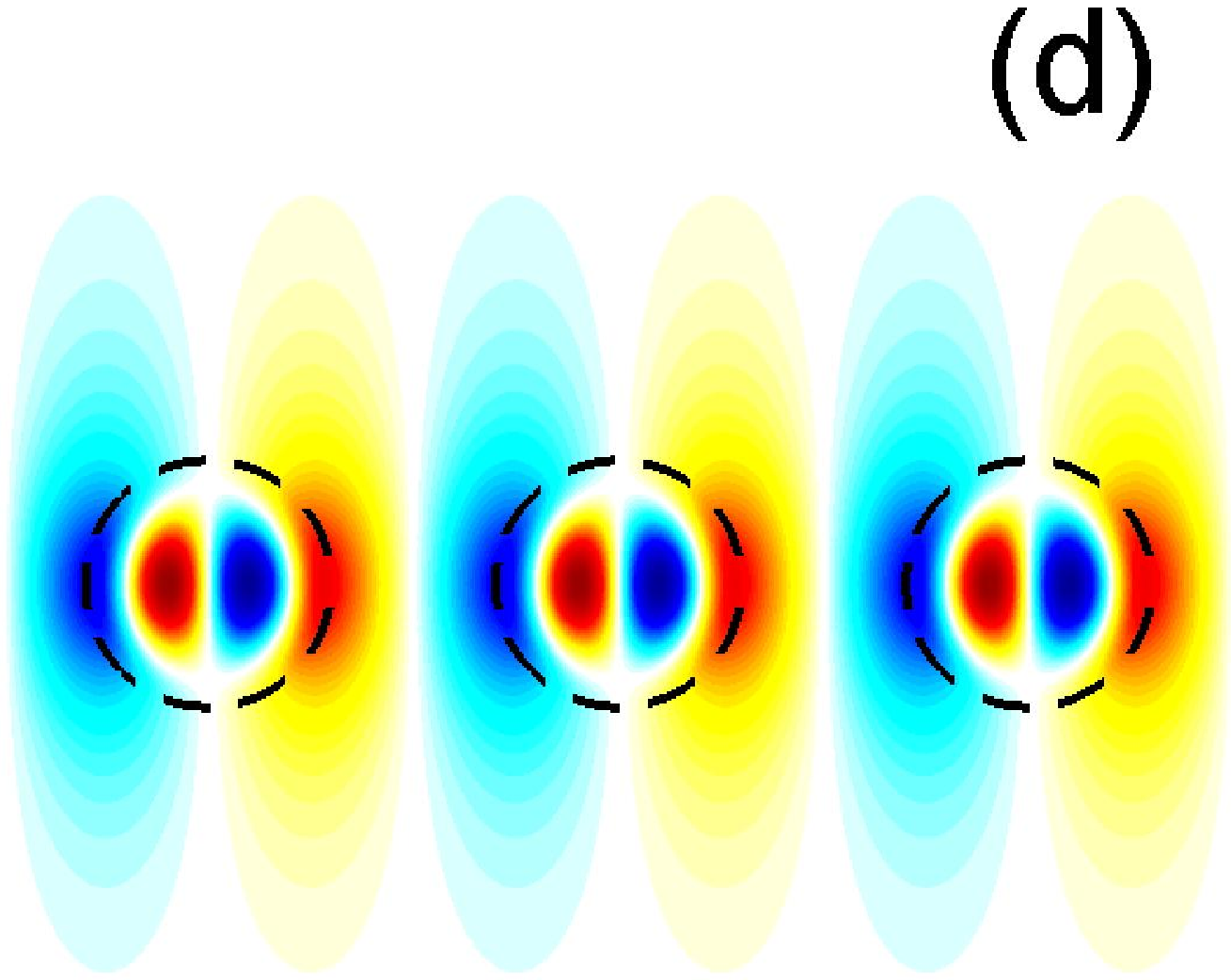}
\caption{(Color online). Patterns of the symmetry protected BSCs which are
solutions of Eq. (\ref{BSC}) for $\epsilon=12, R=0.3$ (a) $k_0=h\omega/c=2.542$,
(b) $k_0=3.6467$, (c) $k_0= 4.85$ and
(d) $k_0= 5.3125$. The array of rods is shown by dash lines.} \label{fig2}
\end{figure}

Let us assume that only diffraction channel $l=0$ is open. Examples of the Bloch states embedded into this
channel are shown in Fig. \ref{fig2} for $k_x=0$
where the pattern of the BSC in Fig. \ref{fig2} (a) was first presented by Shipman and Venakides \cite{Shipman}.
We show that these Bloch BSCs are symmetry protected. For $k_x=0, \pi$
the matrix $\widehat{\Lambda}$ has the property
\begin{equation}\label{Lam1}
    \Lambda_{m,n}=\Lambda_{|m-n|}.
\end{equation}
Let us introduce the operator which inverts the indexes of $a_m$
\begin{equation}\label{P}
    \widehat{P}a_m=a_{-m}.
\end{equation}
Then it follows from Eq. (\ref{Lam1}) that operator $\widehat{P}$ commutes with the matrix
$\widehat{\Lambda}$. Therefore, the eigenvectors of the matrix $\widehat{\Lambda}$ are only symmetric or antisymmetric
relative to $m\rightarrow -m$, in particular the same holds true for the BSC
\begin{equation}\label{BSCsym}
    \Psi_{m,BSC}^{s,a}=\pm\Psi_{-m,BSC}^{s,a}
\end{equation}
where $s$ and $a$ stand for symmetric and antisymmetric states respectively.
For $k_x=0, \pi$ an additional property arises as it could be seen from Eq. (\ref{L})
\begin{equation}\label{Lam2}
    \Lambda_{|2m+1|}=0.
\end{equation}
That allows to split the BSCs into the following
types. The BSCs of the first type have only odd components $a_{2m}^o=0$, while
the BSC of the second type have only even components $a_{2m+1}^e=0$. Thus, the BSCs for
$k_x=0, \pi$  can be split into four types: $(s,o), (s,e), (a,o), (a,e)$.

For diffraction channel $l=0$ with $k_x=0$ we have from Eq. (\ref{vtr})
\begin{equation}\label{vtrl0}
 v_{l=0,m}^{(t)}=\sqrt{\frac{2}{k_{y,0}}}(-1)^m, ~~v_{l=0,m}^{(r)}=\sqrt{\frac{2}{k_{y,0}}}.
\end{equation}
Substituting Eq. (\ref{vtrl0}) into Eq. (\ref{null}) we find that only two types of the BSC
$(a,o)$ and $(a,e)$ are symmetry protected. Let us consider first the Bloch BSC
which belongs to the type $(a,o)$.
Matching the solutions Eq. (\ref{eq6}) on the boundary of the rod we find the BSC wave function interior of the
$j=0$ rod as follows
\begin{equation}\label{odd}
    \Phi(r,\varphi)^{(a,o)}=\sum_{m=1}^{\infty}A_{2m-1}(r)\cos(2m-1)\varphi,
\end{equation}
with
\begin{equation}\label{Am}
    A_m(r)=\frac{\frac{J_m(k_0R)}{T_{mm}}+H_m^{(1)}(k_0R)}{J_m(qR)}a_mJ_m(qr)
\end{equation}
where the property of the Bessel functions $J_{-m}(x)=(-1)^mJ_m(x)$ was taken
into account. One can see that the BSC function has the nodal line at
$\varphi=\pi/2$, i.e., at $x=0$. However because the BSC is a Bloch wave we obtain a periodical set of
nodal lines at $x=0, \pm 1, \pm 2, \ldots$. Note that the BSC exists not only within the rods.
Although there is no contribution of $l=0$ diffraction channel into the scattering function (\ref{scattering}) because
$r_{l=0}=0, t_{l=0}=0$ there are contributions of closed evanescent diffraction channels with $l\neq 0$.
According to  Eq. (\ref{kyl}) the dominant contribution
$l=1$ gives us the localization scale in the y-direction as
\begin{equation}\label{scale}
y_{BSC}\sim 1/\sqrt{4\pi^2-k_0^2}.
\end{equation}
The patterns of the antisymmetric odd Bloch BSC  are shown in Fig. \ref{fig2} (a), (c) and (d).

\begin{figure}[ht]
\includegraphics[height=5cm,width=5cm,clip=]{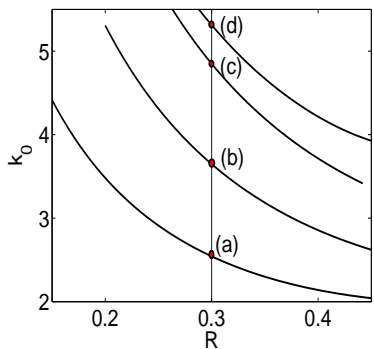}
\caption{BSC frequencies $k_0=\omega/c$ vs radius $R$ for the BSCs shown in Fig. \ref{fig2}.}
\label{fig3}
\end{figure}

For the antisymmetric even BSC $(a,e)$ we have
the following solution within  the rods
\begin{equation}\label{even}
    \Phi(r,\varphi)^{(a,e)}=\sum_{m=1}^{\infty}A_{2m}(r)\sin 2m\varphi,
\end{equation}
which has the nodal lines at $\varphi=0, \pi/2$. The numerically calculated pattern of this BSC is shown in
Fig. \ref{fig2} (b). Thus, the
antisymmetric Bloch BSC exists provided that the BSC frequency $k_0$ does
not exceed the threshold of the next diffraction channels with $l\neq 0$ .
Fig. \ref{fig3} shows that all symmetry protected BSCs from Fig. \ref{fig2} coexist in a wide range of the radius of the rods.

\begin{figure}[ht]
\includegraphics[width=7cm,clip=]{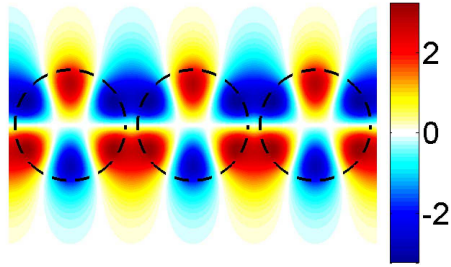}
\caption{(Color online). Real part of the symmetric odd Bloch BSC  for $R=0.4515, k_0=3.0887, k_x=0$.} \label{fig6}
\end{figure}

Next we consider symmetric odd BSC $(s,o)$ in the diffraction channel $k_x=0, l=0$.
Substituting  Eqs. (\ref{vtrl0}) and  (\ref{BSCsym}) into Eq. (\ref{null}) we obtain
\begin{equation}\label{odd sum}
[\mathbf{v}_l^{r,t}]^{+}\mathbf{\Phi}^{(s,o)}=2\sqrt{\frac{2}{k_{y,0}}}\sum_{m=1}^{\infty}\Phi_{2m-1}^{(s,o)}.
\end{equation}
In contrast to the antisymmetric wave the right-hand part of this equation can be equal to zero only accidentally
as one varies the radius of the rods. The corresponding BSC function
\begin{equation}\label{symodd}
    \Phi(r,\varphi)^{(s,o)}=\sum_{m=1}^{\infty}A_{2m-1}(r)\sin (2m-1)\varphi
\end{equation}
is presented in Fig. \ref{fig6}.
This BSC has nodal lines at  $y=0$. The symmetric even BSC $(s,e)$ wave function has the following form for
$r<R$
\begin{equation}\label{symeven}
    \Phi(r,\varphi)^{(s,e)}=\sum_{m=0}^{\infty}A_{2m}(r)\cos 2m\varphi.
\end{equation}
However for given parameter $\epsilon=12$ our computation revealed only the symmetric odd BSC.
Thus, the symmetric BSC both odd and even can not be classified as symmetry protected because of the necessity to
adjust the right-hand part of Eq. (\ref{odd sum}) to zero.
\section{The bound states embedded into two and three diffraction channels}

\begin{figure}[ht]
\includegraphics[width=7cm,clip=]{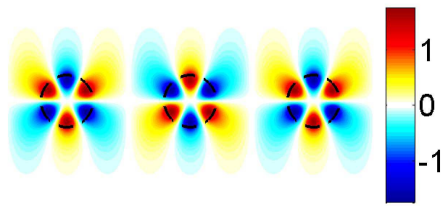}
\caption{(Color online) Real part of the $(s,o)$ BSC wave function embedded into the open $l=0, l=-1$
diffraction channels for $R=0.2107, k_0=6.87, k_x=\pi$.} \label{fig7}
\end{figure}

Now we show that the BSC can exist even when more than one diffraction channel is open.
That phenomenon was considered by Ndangali and Shabanov in the double arrays of dielectric rods \cite{Ndangali}.
They showed that tuning for BSC requires higher dimensionality of the parametric space.
In Ref. \cite{anneal} we demonstrated that the BSC can be robust relative
to a few open channels owe to the symmetry of the resonator. We show in this section that
the present system can also support BSC embedded into a few diffraction channels.
First, we assume that only two diffraction channels $l=0, k_{x,0}=k_x=\pi$  and
$l=-1, k_{x,-1}=-\pi$ are open. The components of ingoing and outgoing waves (\ref{vtr}) satisfy the following relationships
\begin{equation}\label{vtrl-1}
 v_{l=0,m}^{(r,t)}=[v_{l=-1,m}^{(r,t)}]^{*}=v_{l=-1,-m}^{(r,t)}.
\end{equation}
In this case all symmetric properties established in Eqs.  (\ref{BSCsym}) and (\ref{Lam2}) still hold true. We start with the symmetric
odd state $(s,o)$ and consider its coupling with the channels (\ref{vtrl-1}) to establish the following equalities
\begin{equation}\label{Wl--1}
W_{l=0}^{(r,t)}=W_{l=-1}^{(r,t)}, ~~W_{l=0}^{(r)}=-W_{l=0}^{(t)}.
\end{equation}
Here, for brevity we omitted the superscript $(s,o)$. One can see that all coupling constants equal zero if $W_{l=0}^{(t)}=0$
which can be fulfilled through variation of the radius of the rod. Such  Bloch BSC function
is shown in Fig. \ref{fig7}.
For $\epsilon=12$ we did not find the BSC of other symmetry types:
 $(s,e), (a,o), (a,e)$. Although that does not mean that they are impossible for different $\epsilon$.

Next, we consider  the case of three open diffraction channels
$l=0, k_{x,0}=k_x=\pi$, $l=1, k_{x,1}=\pi$, and $l=-1, k_{x,-1}=-\pi$, the utmost number of open channels
with an embedded bound state. Our simulations show that the only antisymmetric odd BSC $(a,o)$ can exist for which the symmetry
establishes five equalities for six coupling constants
\begin{equation}\label{Wl-1}
W_{l=-1}^{(r,t)}=-W_{l=1}^{(r,t)}, ~~W_{l=1}^{(r)}=W_{l=1}^{(t)}, ~~
W_{l=0}^{(r)}=W_{l=0}^{(t)}=0.
\end{equation}
Similar to the previous case in order to cancel all couplings it is enough to fulfill
$W_{l=1}^{(t)}=0$. The last equality can be achieved by
tuning the radius of the rods. The wave function within the rods has the following form
\begin{equation}\label{anti odd}
    \Phi(r,\varphi)^{(a,o)}=\sum_{m=1}^{\infty}A_{2m-1}(r)\cos (2m-1)\varphi.
\end{equation}
\begin{figure}[ht]
\includegraphics[width=7cm,clip=]{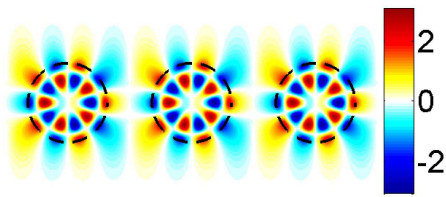}
\caption{(Color online) Real part of the $(a,o)$ BSC wave function embedded into the three open diffractions
channels: $l=0, l=\pm 1$ for $ R=0.3193, k_0=9.876, k_x=0$.} \label{fig8}
\end{figure}
The Bloch wave function is shown in Fig. \ref{fig8}. One can see that nodal lines at $x=0, \pm 1/2, \pm 1, \ldots$ are
similar to the antisymmetric odd Bloch BSC embedded in the first diffraction channel shown in
Fig. \ref{fig2}.

\section{Bloch BSC with Poynting vector}
Could the Bloch BSC occur at $\pi>k_x > 0$ in the continuum of free-space modes?
This question was first answered positively by Porter and Evans \cite{Porter} who
considered acoustic trapping in the array of rods of rectangular cross-section.
Later Marinica {\it et al} \cite{Shabanov} demonstrated the existence of the Bloch BSC at $k_x > 0$
in two parallel dielectric gratings and Ndangali and Shabanov \cite{Ndangali} in two
parallel arrays of dielectric rods. Each array has a transmission zero for definite incident angle and
frequency to form the Fabry-Perot resonator that support BSCs trapped between the "mirrors" \cite{Shabanov}.
In a single array of rods positioned on the surface of bulk 2d photonic crystal multiple BSCs with $k_x\geq 0$ were
considered by Wei {\it et al} \cite{Wei1}.
That system can be also seen as a Fabry-Perot resonator with
the photonic crystal playing the role of  bottom mirror provided that the frequency is in the band gap.
The periodic array of rods plays the role of top "mirror" at certain  frequency and the radius of the rods.
Such BSCs are one-dimensional Bloch surface states which do not leak into the radiation continuum \cite{Kivshar}.
The next important step was undertaken by Zhen {\it et al} \cite{Zhen} who presented the Bloch BSC
in a PhC slab with one-dimensional periodicity in $x$ where the Bloch BSC evolves with
$k_x\neq 0, k_z=0$ into
the BSC with $k_x=0, k_z\neq 0$ for decreasing of the slab thickness.

\begin{figure}[ht]
\includegraphics[width=5cm,clip=]{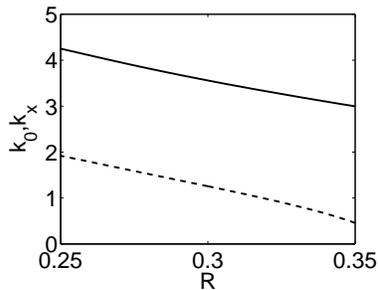}
\caption{$k_x$ (solid line) and $k_0$ (dash line) vs radius of rods $R$.}
\label{fig4}
\end{figure}

\begin{figure}[ht]
\includegraphics[width=12cm,clip=]{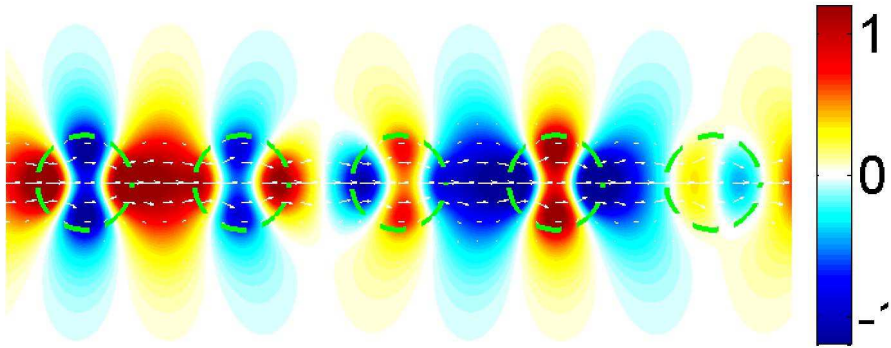}
\caption{(Color online) Real part of the propagating BSC  for $R=0.3, k_0=3.5577, k_x=1.2556$. White arrows
show the Poynting vector carrying power current in the Bloch BSC.} \label{fig5}
\end{figure}

In the present paper we show that the electromagnetic Bloch BSC with $\pi> k_x > 0$ occurs in a periodic array of
rods of the circular cross-section as the result  of zero coupling of the Bloch state
with diffraction channel  at specific values of $k_x$ and $k_0$.
We stress that similar to the systems described above \cite{Wei,Wei1,Yang} there is no necessity to tune the material
parameters of the rods. The only condition is that the dielectric constant and the radius of the rods are
large enough for only $l=0$ diffraction channel to be open as seen from Fig. \ref{fig4}.
These Bloch BSCs are degenerate due to $k_x\rightarrow -k_x$ symmetry. Fig. \ref{fig5} shows the pattern of the BSC
(real part) with the Poynting vector field that indicates a power flux along the array.
The diffraction channel parameters can be changed as one changes the radius of the rods
as shown in Fig. \ref{fig4}.
\section{Emergence of the BSC in scattering function}
It is clear that probing the BSC by a wave injected into a diffraction channel is impossible because the BSC is orthogonal to
the channel. Nevertheless one can show there is a path in the parametric space leading to the BSC point.
Approaching the BSC point along this path reveals the BSC as dominant contribution in the scattering wave function (\ref{scat}).
Assume that among all complex eigenvalues $\lambda_f$ only one tends to zero $\lambda_s\rightarrow 0$ while the others
remain finite  in the vicinity of the BSC point. In what follows we suppose that only one
 diffraction channel $l=0$ is open. Then we can write the singular part in the  scattering wave (\ref{scat}) as follows
\begin{equation}\label{BSCscat}
    \mathbf{a}_s=\frac{(\mathbf{y}_s^{+}\mathbf{v}_0^{(t,r)})\mathbf{x}_s}{\lambda_s}
\end{equation}
where $\mathbf{x}_s$ is the eigenvector which tends to the null vector.
Respectively,
we can present the singular parts in the reflection and transmission amplitudes (\ref{rt}), i.e.,
the scattering matrix as follows
\begin{equation}\label{rsts}
    r_{sl}=\frac{W_{s,l}^{(r)}\widetilde{W}_{s,0}^{(t)}}{\lambda_s}, ~~
    t_{sl}=\frac{W_{s,l}^{(t)}\widetilde{W}_{s,0}^{(t)}}{\lambda_s}.
\end{equation}

\begin{figure}[ht]
\includegraphics[width=10cm,clip=]{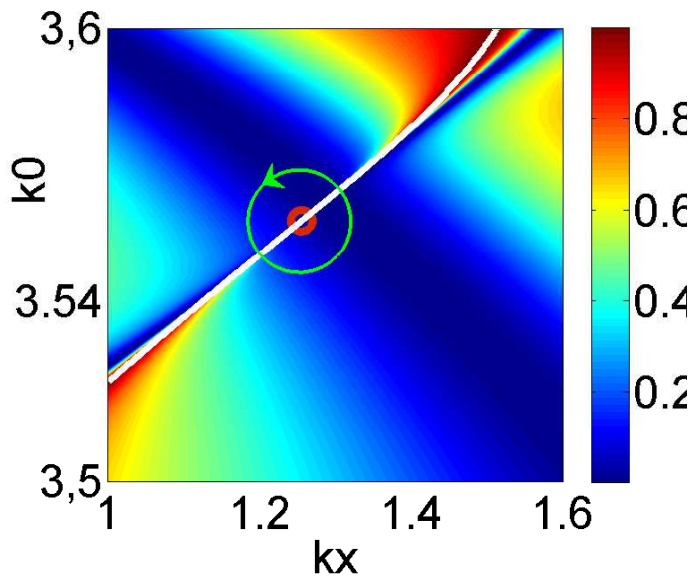}
\caption{(Color online) Transmittance vs $k_0$ and $k_x$ for a wave incident to the array of dielectric rods
 shown in Fig. \ref{fig1}.
 The BSC point $k_0=3.5577, k_x=1.2556, R=0.3$ is shown by red open circle.} \label{fig9}
\end{figure}

Fig. \ref{fig9} shows the transmittance vs $k_x$ and $k_0$.
One can see a singular point where unit transmittance touches zero transmittance which corresponds to
the collapse of the Fano resonance \cite{Kim}. This singular point $k_{xs}, k_{0s}$ corresponds to the Bloch BSC shown in
Fig. \ref{fig5}.
To find the path leading to the BSC like scattering wave
we following Ref. \cite{ring} consider the behavior of the scattering function as one encircles the singular point as
shown in Fig. \ref{fig9} by green arrow.
\begin{figure}[ht]
\includegraphics[height=5cm,width=5cm,clip=]{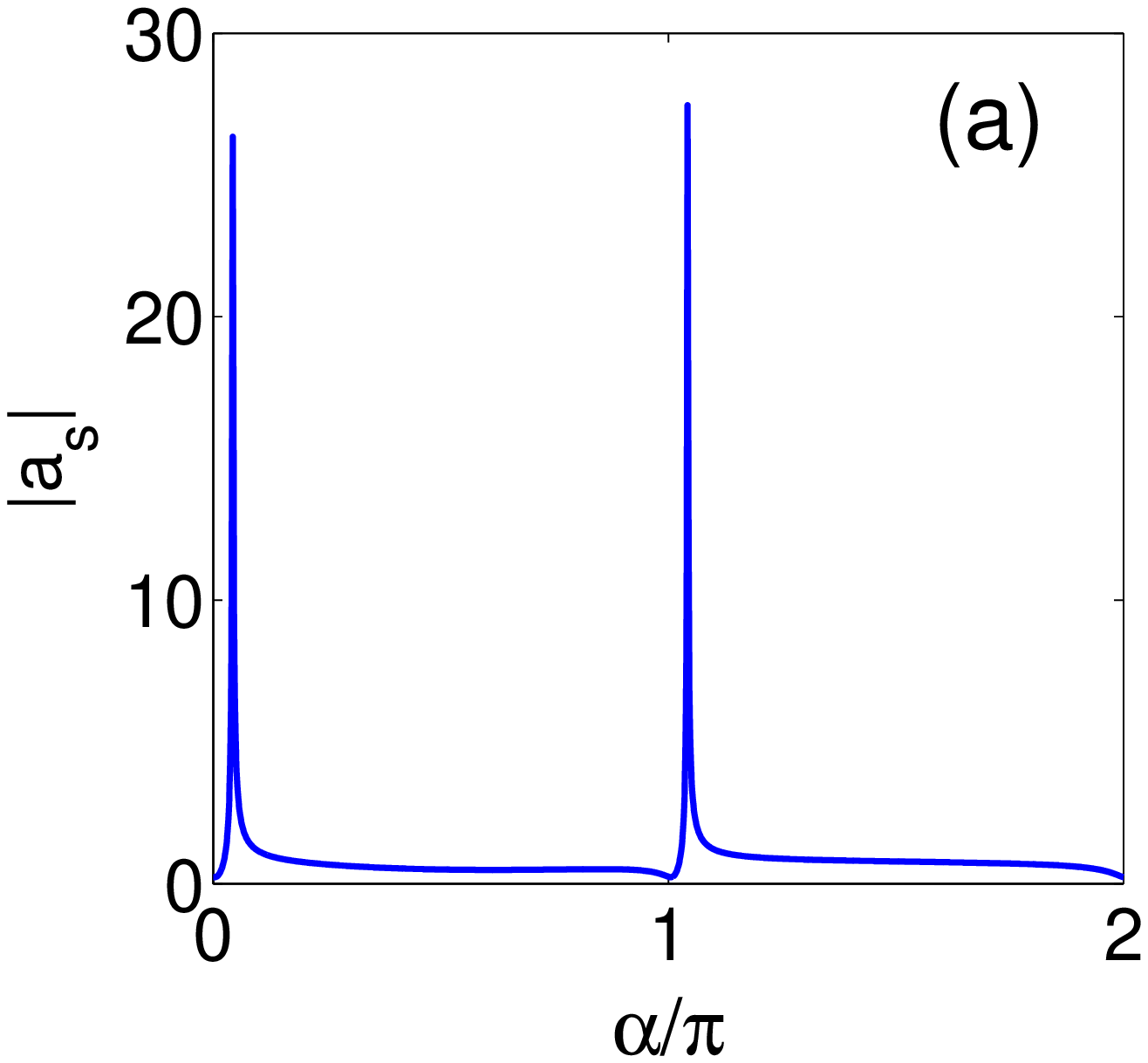}
\includegraphics[height=5cm,width=5cm,clip=]{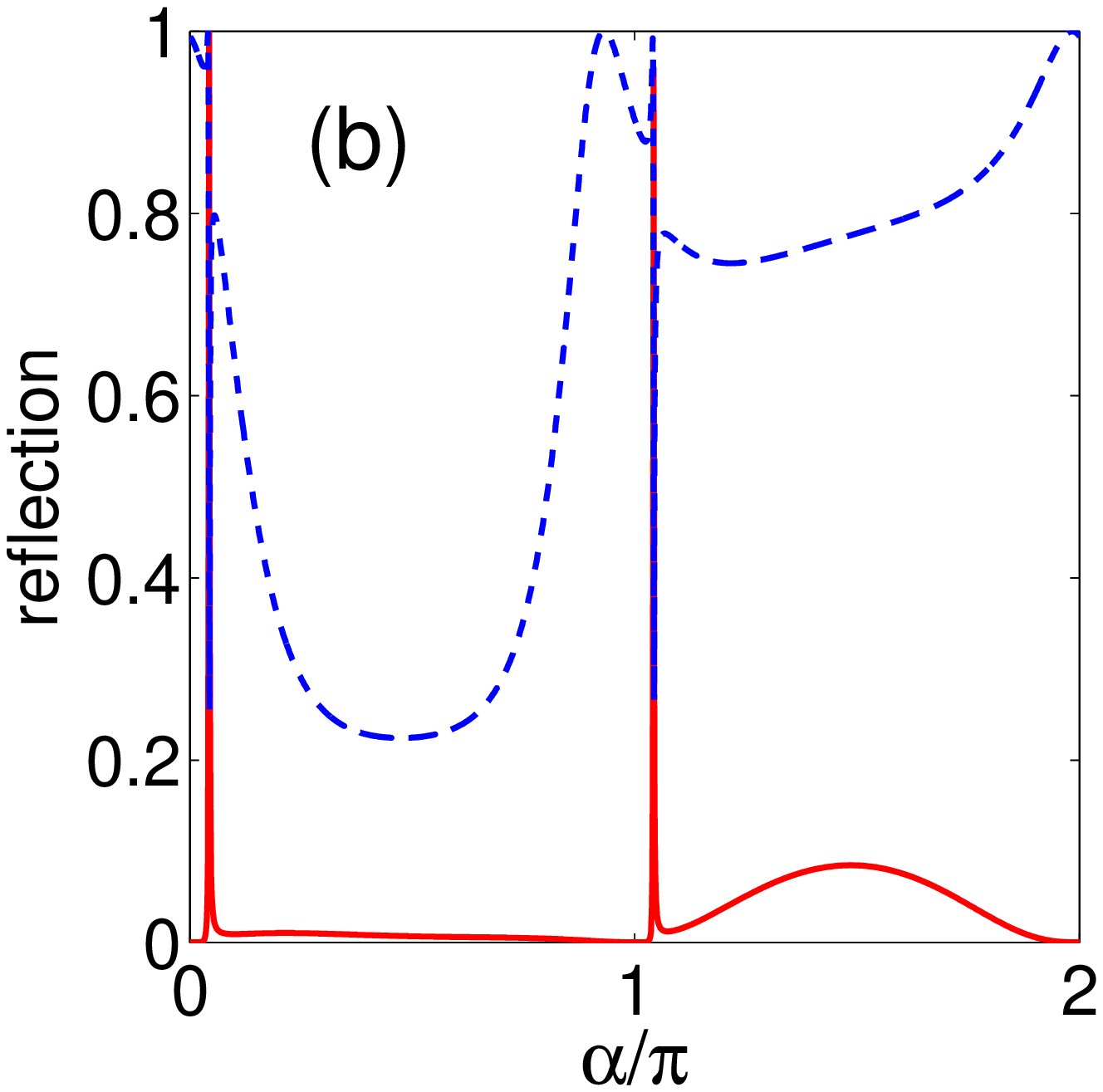}
\caption{(Color online) Behavior of $|a_s|$ (a) and reflection coefficient (b) vs angular variable $\alpha$ for encircling of the BSC point
shown in Fig. \ref{fig9}.} \label{fig10}
\end{figure}
In Fig. \ref{fig10} we plot the singular part in the  scattering state (\ref{BSCscat}) $|a_s|$ and reflection coefficient
$|r_{sl=0}|$  as a function of the angle variable $\alpha$ defined via
$k_x-k_{xs}=\rho\cos\alpha, ~k_0-k_{0s}=\rho\sin\alpha$ with $\rho=0.1$.  The angular behavior demonstrates
sharp peaks. The smaller is the radius $\rho$ the sharper are the peaks associated with the BSC.
Therefore it follows that there is a path leading to the BSC-like scattering function.
This path is shown by white solid line
in Fig. \ref{fig9} which corresponds to the unit transmittance in the collapsing Fano resonance.
\section{Summary}

In the present paper we used the approach presented in Ref.
\cite{Yasumoto} for cylindrical rods. The approach is based on expansion of incident and scattered waves over the
Hankel or Bessel functions and has an advantage that the number of functions could be taken rather
small. The symmetry of the system waves implies the symmetry selection rules for
coefficients $a_m$ in the expansion of the Bloch BSC. That results in four types for the Bloch BSC with wave vector $k_x=0,\pi$
classified in Section IV.
In turn the symmetry restrictions for $a_m$ are reflected in the space symmetry as demonstrated in Fig. \ref{fig2}.
It is important the symmetry protected Bloch BSC exist in a wide range of the material parameters
of the rods. This symmetry property allows the existence of the Bloch BSC embedded into two and three
diffraction channels  although at the price of tuning the radius of the rods. Such BSCs are presented
in Figs. \ref{fig6}-\ref{fig8}.
The next interesting class of the Bloch BSC is the BSC  with nonzero Bloch wave vector show in Fog. \ref{fig5}.
This BSC carries a power flux along the array.

The BSC exist in a selected point in the parametric space. So it might be thought that the BSC are
not important beyond this point. Following Ref. \cite{ring} we show that the scattering wave function
becomes BSC-like if one approaches to the BSC point alone the path where the
transmission coefficient equals unit (see Figs. \ref{fig9} and \ref{fig10}). This
provides a cue to experimental observation of the BSC.

{\bf Acknowledgments}.
 The work was supported by Russian Science Foundation through the grant 14-12-00266.
 We acknowledge discussions with  D.N. Maksimov.

\end{document}